\documentclass{iaus}
\usepackage{fancybox}

\title[Obtaining planetary parameters using MC analysis] 
{Determination of stellar, orbital and planetary 
        parameters using complete Monte-Carlo analysis -- the case 
        of HAT-P-7b}

\author[P\'al et al.]   
{Andr\'as P\'al$^{1,2}$, 
G\'asp\'ar \'A. Bakos$^1$,
Robert W. Noyes$^1$
\and Guillermo Torres$^1$}

\affiliation{%
$^1$Harvard-Smithsonian Center for Astrophysics, \\ 
	60 Garden st., Cambridge, MA 02138, USA \\ 
	email: {\tt apal@cfa.harvard.edu} \\[\affilskip]
$^2$Dept. of Astronomy, E\"otv\"os Lor\'and University, \\ 
	P\'azm\'any Peter stny 1/A, Budapest, 1117, Hungary \\
}

\pubyear{2008}				
\volume{253}  				
\pagerange{137-140}			
\jname{Transiting Planets}		
\editors{F. Pont}			
\begin{document}

\maketitle

\begin{abstract}
The recently discovered transiting very hot Jupiter, 
HAT-P-7b, a planet
detected by the telescopes of HATNet, turned out to be among the ones
subjected to the highest irradiation from the parent star. As 
known, the combination of photometric and
spectroscopic data for such an object yields the stellar, orbital and
planetary parameters. In order to best characterize this particular
planet, we carried out a complex analysis based on a complete and
simultaneous Monte-Carlo solution using all available data. We included
the discovery light curves, partial follow-up light curves, the
radial velocity data, and we used the stellar evolution models to infer
the stellar properties. 

This self-consistent way of modeling provides the most precise estimate of
the a posteriori distributions of all of the system parameters of interest,
and avoids making assumptions on the values and uncertainties of any of the
internally derived variables describing the system. 
This analysis demonstrates that even partial light curve
information can be valuable. This may become very important for
future discoveries of planets with longer periods -- and therefore
longer transit durations -- where the chance of observing a full event is
small.
\end{abstract}

\firstsection 

\section{Introduction}

The transiting extrasolar planet candidate HTR154-011 has been identified 
by the telescopes of the HATNet (HAT-7, HAT-8 and HAT-6, HAT-9) --
see \cite{bakos2002} for a detailed description about the instrumentation. 
The reduction of the frames followed 
the standard frame calibration procedures, followed by star detection, 
astrometry and aperture photometry which was supported 
by various detrending algorithms.
We search the light curves for 
box-shaped transit signals using the BLS algorithm (\cite{kovacs2002}): 
a periodic dip was detected in the light
curve of 2MASS~19285935+4758102.
The initial ``rejection mode'' spectroscopy with CfA Digital
Speedometer (\cite{latham1992}) at FLWO eliminated the possibility
of an eclipsing binary star system. The analysis of Keck HIRES
spectra resulted the radial velocity data, stellar atmospheric parameters 
and the bisector information (which ruled out a 
``false positive'' hierarchical triple).
We also carried out a partial photometry follow-up in the Sloan $z$-band
with KeplerCam on the FLWO~1.2m telescope. See \cite{pal2008} for more
details about the observational details and data reduction.

The basics of the characterization of transiting extrasolar planets
are the following. The period of the orbital motion can be very 
accurately determined because of the long timespan of the discovery 
(HATNet) observations. The mass of the planet is proportional to the amplitude
of the RV curve, $K$. The radius of the planet relative to the star determines 
the depth of the transit (which is resulted both by the follow-up light curves
and the HATNet data). The inclination of the orbit can be derived from 
the shape of the transit light curve, i.e. the impact parameter is
proportional to the cosine of the inclination. To calculate the absolute 
mass and radius, we need the same data or the host star too, therefore we 
need at least one luminosity indicator for a good stellar evolution model. 
Because of the lack of parallax, this indicator can be $a/R_\star$, 
see \cite{sozzetti2007}. The bottleneck in this case is the partial follow-up,
i.e. the shape parameters of the light curve -- namely the 
ingress/egress length relative to the total duration -- cannot 
be derived directly to obtain a reasonable value of $a/R_\star$.

\section{System parameters}

To obtain the planetary, stellar and orbital parameters, we 
used all available data: 
(i) The HATNet light curves with 
approximately $16,600$ data points 
(moderate precision with an rms of 7\,mmag). The HATNet data
are affected by the period $P$, epoch $T_{\rm center}$ (strongly),
the planetary radius $p$ (moderately), duration of the 
transit $T_{\rm dur}=2(\zeta/R_\star)^{-1}$ (moderately), and the 
impact parameter $b$ (slightly).
(ii) Radial velocity data, high precision relative to the amplitude.
The RV data are affected strongly by the period $P$, 
epoch $T_{\rm center}$, semi-amplitude $K$, and eccentricity 
$k=e\cos\omega$, $h=e\sin\omega$.
(iii) The partial follow-up light curve, acquired shortly after the 
radial velocity measurements. The follow-up photometry data are
affected by the period $P$ (slightly), epoch $T_{\rm center}$
(strongly), planetary radius $p$ (strongly), the duration of the transit 
$T_{\rm dur}$ (strongly), and the impact parameter $b$ (strongly).

Because there was only a couple of periods time difference between
the radial velocity observations and the partial follow-up photometry, 
the transit center time could have been very well extrapolated,
thus the total transit duration can also be accurately obtained.
There is a strong coupling between the adjustable parameters,
moreover, the contribution of each parameter in each type
of data are different (see above), therefore to obtain a plausible
set of system parameters, a joint fit should be performed, incorporating
all available data simultaneously.

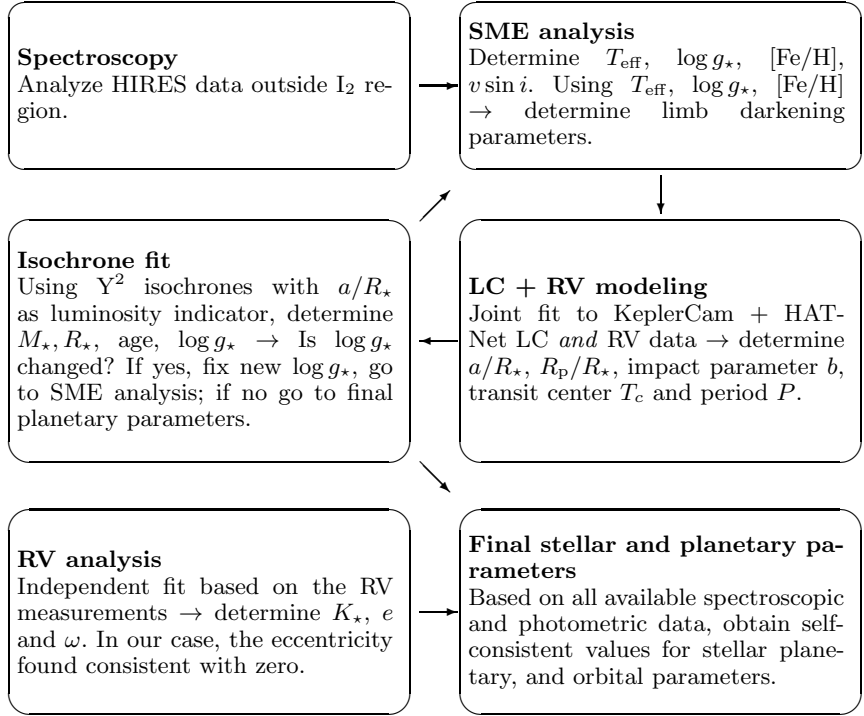
\begin{figure}
\begin{center}
\setlength{\unitlength}{1mm}
\begin{picture}(96,100)
\put(20,86){\makebox(0,0){\cornersize{0.2}\ovalbox{%
\begin{minipage}[c][20mm]{50mm}
\small
\textbf{Spectroscopy}

Analyze HIRES data outside $\mathrm{I}_2$ region.
\end{minipage}
}}}
\put(48,86){\vector(1,0){5}}
\put(80,86){\makebox(0,0){\cornersize{0.2}\ovalbox{%
\begin{minipage}[c][20mm]{50mm}
\small
\textbf{SME analysis}

Determine $T_{\mathrm{eff}}$, $\log g_\star$, $[\mathrm{Fe}/\mathrm{H}]$, 
$v\sin i$. Using $T_{\mathrm{eff}}$, $\log g_\star$, $[\mathrm{Fe}/\mathrm{H}]$
$\to$ determine limb darkening parameters.
\end{minipage}
}}}
\put(80,74){\vector(0,-1){5}}
\put(20,52){\makebox(0,0){\cornersize{0.2}\ovalbox{%
\begin{minipage}[c][30mm]{50mm}
\small
\textbf{Isochrone fit}

Using Y$^2$ isochrones with $a/R_\star$ as luminosity indicator,
determine $M_\star, R_\star$, age, $\log g_\star$
$\to$
Is $\log g_\star$ changed? If yes, fix new $\log g_\star$, go to SME
analysis; if no go to final planetary parameters. 
\end{minipage}
}}}
\put(48,68){\vector(1,1){4}}
\put(53,52){\vector(-1,0){5}}
\put(80,52){\makebox(0,0){\cornersize{0.2}\ovalbox{%
\begin{minipage}[c][30mm]{50mm}
\small
\textbf{LC + RV modeling}

Joint fit to KeplerCam + HATNet LC \emph{and} RV data $\to$
determine $a/R_\star$, $R_{\mathrm{p}}/R_\star$, impact parameter $b$, transit
center $T_c$ and period $P$.
\end{minipage}
}}}
\put(48,36){\vector(1,-1){4}}
\put(20,16){\makebox(0,0){\cornersize{0.2}\ovalbox{%
\begin{minipage}[c][25mm]{50mm}
\small
\textbf{RV analysis}

Independent fit based on the RV measurements $\to$ determine $K_\star$,
$e$ and $\omega$. 
In our case, the eccentricity found consistent with zero.
\end{minipage}
}}}
\put(48,16){\vector(1,0){5}}
\put(80,16){\makebox(0,0){\cornersize{0.2}\ovalbox{%
\begin{minipage}[c][25mm]{50mm}
\small
\textbf{Final stellar and planetary parameters}

Based on all available spectroscopic and photometric data, obtain
self-consistent values for stellar planetary, and orbital parameters.
\end{minipage}
}}}
\end{picture}
\end{center}
\caption{Flowchart of the iterative Monte-Carlo analysis 
of the stellar, orbital and planetary parameters. The arrows indicate
the ``data flow'' between the subsequent steps. See text for further details.}
\end{figure}

\section{Monte-Carlo modeling}

The basic system parameters (see previous section) can be determined 
using a Monte-Carlo algorithm (the method of refitting to 
synthetic data sets, MCMC), resulting an \emph{a posteriori} distribution 
of the adjustable parameters. This method also automatically 
results the \emph{a posteriori} distribution of derived quantities, such 
as $a/R_\star$ or the inclination.
The luminosity indicator is the density of the star $\rho_\star$ which is
related to the observable quantities as
\begin{equation}
\rho_\star = \frac{3\pi}{GP^2}\left(\frac{a}{R_\star}\right)^3-
\frac{3K}{2PG\sin i}\left(\frac{a}{R_\star}\right)^2\frac{1}{R_\star}
\equiv \rho_0 - \frac{\Sigma_0}{R_\star}, \label{rhostar}
\end{equation}
where both $\rho_0$ and $\Sigma_0$ are observables, therefore these also have 
an \emph{a posteriori} distribution resulted by the Monte-Carlo fit.
The stellar evolution model (\cite{yi2001}) results the radius of the 
star as the function of the density, surface temperature $T$ and
metallicity $z$, i.e. 
\begin{equation}
R_\star=R_\star(\rho_\star,T,z). \label{rstar}
\end{equation}
Equation~(\ref{rhostar}) and equation~(\ref{rstar}) have two unknowns, 
$R_\star$ and $\rho_\star$ since $T$ and $z$ are also observables, 
known from the spectroscopy (and also have a MC 
distribution from the SME analysis). 
Therefore, this set of equations can be solved, which yields
the \emph{a posteriori} distribution of the stellar parameters.
The combination of the stellar parameters $R_\star$, $M_\star$ with
the previously fitted quantities results the \emph{a posteriori} 
distribution of the planetary parameters in a straightforward way.
A sanity check can be done by comparing the resulted stellar surface gravity
(obtained directly from $M_\star$ and $R_\star$) with the 
value provided by the SME. If they differ, the SME analysis has to
be redone by fixing the surface gravity and
the whole procedure can be done repeatedly until convergence,
see flowchart on Fig.~1. We note that the complete MC way of analysis results
the derived parameters independently from the form of calculations, e.g.
the \emph{a posteriori} distribution of the semimajor axis is 
exactly the same if it is calculated as $a=R_\star(a/R_\star)$ or 
$a=[G(M_\star+M_{\rm p})]^{1/3}[P/(2\pi)]^{2/3}$, 
altough $R_\star$ or $a/R_\star$
themselves have definitely larger individual uncertainties than 
the period or $M_\star^{1/3}$. 
The final stellar, planetary and orbital
parameters for the HAT-P-7(b) system are summarized in Table~1.

\begin{table}[!ht]
\caption{Stellar, orbital and planetary parameters of the HAT-P-7(b)
system.}
\begin{center}\begin{tabular}{lcl}
\hline
{\bf Stellar parameter} & {\bf Value} & {\bf Source} \\
\hline
$T_{\mathrm{eff}}$ ($\mathrm{K})$       & $6350\pm80$           & SME \\
$[\mathrm{Fe}/\mathrm{H}]$              & $+0.26\pm0.08$        & SME \\
$\log g_*$ (cgs)                        & $4.07\pm0.06$         & Y$^2$+LC+SME \\
$v\sin i$ ($\mathrm{km~s^{-1}}$)        & $3.8\pm0.5$           & SME \\
$M_*$ ($M_\odot$)                       & $1.47\pm0.06$         & Y$^2$+LC+SME \\
$R_*$ ($R_\odot$)                       & $1.84^{+0.23}_{-0.11}$& Y$^2$+LC+SME \\
$M_V$ (magnitude)                       & $3.00\pm0.22$         & Y$^2$+LC+SME \\
Distance (pc)                           & $320^{+50}_{-40}$     & Y$^2$+LC+SME \\
\hline 
\end{tabular}\vspace*{5mm}

\begin{tabular}{lc}
\hline
{\bf Orbital/planetary parameter} & {\bf Value} \\
\hline
$P$ (days)                                      & $2.204730\pm0.000004$         \\
$T_{\mathrm{center}}$ (BJD)                     & $2,453,790.2593\pm0.0010$     \\
$a/R_*$                                         & $4.35^{+0.28}_{-0.38}$        \\
$p\equiv R_{\mathrm{p}}/R_*$                    & $0.0763 \pm 0.0010$           \\
$b\equiv a\cos i/R_*$                           & $0.37^{+0.15}_{-0.29}$        \\
$i$ (degrees)                                   & $85.^\circ7 ^{+3.5}_{-3.1}$   \\
\hline
$K$ ($\mathrm{m~s^{-1}}$)                       & $213.5\pm1.9$                 \\
$e$                                             & $0.003\pm0.012$               \\
\hline
$M_{\mathrm{p}}$ ($M_{\mathrm{J}}$)             & $1.776^{+0.077}_{-0.049}$     \\
$R_{\mathrm{p}}$ ($R_{\mathrm{J}}$)             & $1.363^{+0.195}_{-0.087}$     \\
$a$ (AU)                                        & $0.0377\pm0.0005$             \\
\hline
\end{tabular}\end{center}

\end{table}

\section{Summary}
The outlined method provides a straightforward way
for characterizing planetary systems since 
	(i) it omits any kind of intermediate values which otherwise
        would distort the error estimations,
        (ii) automatically results the uncertainties and correlations
        of the system parameters, including the stellar, planetary
        and orbital parameters too,
        (iii) it can utilize even the partial light curves which 
        otherwise would not provide enough information for 
        further steps  --
        these partial follow-up light curves become even more important
        when systems with longer periods will be discovered (i.e.
        the chance to observe a complete transit on a single night
        will be definitely smaller).
The planet HAT-P-7b is also very interesting since
	(i) it has a very high surface temperature, 
        somewhere between $2200-2700$~K, depending on the heat redistribution,
        (ii) it happens to fall in the field of view of the upcoming Kepler 
        Mission, yielding an opportunity for further continuous observations.

\section*{Acknowledgments}

The work of A.~P. was supported by the NASA grant NNG04GN74G. Work
by G.~\'A.~B. was supported through Hubble Fellowship Grant 
HST-HF-01170.01-A and by the NSF postdoctoral fellowship grant AST-0702843. 
Operation of the HATNet is supported by the NASA grants 
NNG04GN74G and NNX08AF23G.

{}


\begin{thebibliography}{}

\bibitem[Bakos et al.(2002)]{bakos2002}
Bakos, G. \'A., L\'az\'ar, J., Papp, I., S\'ari, P., Green, E. M.,
2002, \textit{PASP}, 114, 974

\bibitem[Kov\'acs, Zucker, \& Mazeh(2002)]{kovacs2002}
Kov\'acs, G., Zucker, S. \& Mazeh T.,
2002, \textit{A\&A}, 391, 36

\bibitem[Latham(1992)]{latham1992}
Latham, D. W.,
1992, IAU Coll. 135; \textit{ASP Conf. Ser.} 32, 110

\bibitem[P\'al et al.(2008)]{pal2008}
P\'al, A. et al.,
2008, \textit{ApJ}, 680, 1450

\bibitem[Sozzetti et al.(2007)]{sozzetti2007}
Sozzetti, A. et al.,
2007, \textit{ApJ}, 664, 1190

\bibitem[Yi et al.(2001)]{yi2001}
Yi, S. K. et al., 
2001, \textit{ApJS} 136, 417

\end{thebibliography}
\end{document}